\documentclass{iopconfser}

\begin{document}

\title{On the stability of de Sitter inflationary solution in the Starobinsky-Bel-Robinson gravity}

\author{Tuan Q. Do$^{1}$}

\affil{$^1$Phenikaa Institute for Advanced Study, Phenikaa University, Hanoi 12116, Vietnam}
\email{tuan.doquoc@phenikaa-uni.edu.vn}

\begin{abstract}
We will present the way to derive a de Sitter inflationary solution within the so-called Starobinsky-Bel-Robinson gravity. Then, we will show by using the dynamical system method whether the obtained solution is stable or not. According to the stability of the de Sitter inflationary solution, we could judge which phase of our universe, among the two early and late-time phases, is more appropriate for this solution. 
\end{abstract}

\section{Introduction}
Inflationary universe \cite{Starobinsky,Guth,Linde} has played a central role in modern cosmology for recent decades. Remarkably, the Starobinsky gravity model \cite{Starobinsky} has been widely considered as one of the most successful inflationary models in the light of the latest Planck data \cite{Planck}. 
As a result, the Starobinsky gravity model's action is given by  
\begin{equation}
		S =\frac{M_p^2}{2} \int d^4 x  \sqrt{-g}  \left[R+\beta_1 R^2\right] ,
\end{equation}
where $M_p$ is the reduced Planck mass, while $R^2$ is nothing but the second-order term of the Ricci scalar, which arises due to quantum corrections as pointed out by Starobinsky in Ref. \cite{Starobinsky}. Remarkably, the Starobinsky gravity model does not involve an inflaton, which is a scalar field hypothesized to cause an inflationary phase of the  early universe \cite{Linde}. However, it can be conformally transformed to an effective model of inflaton \cite{Whitt:1984pd}. One can therefore obtain a mass of this inflaton in terms of a relation as $\beta_1 = 1/(6m^2)>0$. Several key points of the Starobinsky gravity model can be listed for interested readers as follows: (i)
It does not involve any exact de Sitter inflationary solution, but a quasi-de Sitter one; (ii) it seems to be the only higher-order gravity model being free from the so-called Ostrogradsky instability \cite{Woodard:2015zca};  (iii) it belongs to higher-order curvature gravity models, which are potential candidates for revealing quantum gravity \cite{Stelle:1976gc}; and (iv) it belongs to $f(R)$ gravities, which are relevant to the late-time accelerated expansion of our universe \cite{Nojiri:2010wj}. Despite of these advantages of the Starobinsky model, however, the precise CMB measurements would require slight modifications of its predictions, which might only be derived from its non-trivial extensions. Indeed, the Starobinsky has been extended widely by many people via various approaches. Currently, we are interested in the so-called Starobinsky-Bel-Robinson (SBR) gravity model, which has been proposed by Ketov in Ref. \cite{Ketov} and discussed extensively in Refs. \cite{Ketov1,Ketov2}. In particular, we would like to examine whether a stable de Sitter inflationary solution can exist within this model. As a result, we have been able to derive an exact de Sitter inflationary to the SBR model. However, stability analysis has been performed to show that this solution is indeed unstable, consistent with the previous study made in Ref. \cite{Ketov1}.  Detailed results will be presented in the next sections. 

It can be stated that this proceedings article is nothing but a short summary of isotropic part of our recent study on the SBR gravity model published in Ref. \cite{Do:2023yvg}. It will be organized as follows: (i) A short introduction of our study has been written in the section 1. (ii) Basic setup of the SBR gravity model along with its de Sitter inflationary solution will be presented in the section 2. (iii) The stability issue of the obtained de Sitter solution will be discussed in the section 3. (iv) Finally, conclusions will be given in the section 4.
\section{Starobinsky-Bel-Robinson gravity model}
\subsection{Action}
An action of the SBR gravity model considered by Ketov in Ref. \cite{Ketov} takes the following form,
\begin{equation}
S_{\rm SBR}= \frac{M_p^2}{2}\int d^4 x \sqrt{-g}  \left[R+\frac{1}{6m^2} R^2 -\frac{\beta}{8m^6}T^{\mu\nu\lambda\rho}T_{\mu\nu\lambda\rho} \right] ,
\end{equation}
where $T^{\mu\nu\lambda\rho}$ is the four-dimensional Bel-Robinson (BR) tensor, whose definition is given by \cite{Bel,Robinson1,Robinson2}
\begin{equation}
T^{\mu\nu\lambda\rho} =R^{\mu\alpha\beta\lambda} R^\nu{}_{\alpha\beta}{}^\rho +R^{\mu\alpha\beta\rho}R^\nu{}_{\alpha\beta}{}^\lambda-\frac{1}{2}g^{\mu\nu}R^{\alpha\beta\gamma\lambda}R_{\alpha\beta\gamma}{}^\rho.
\end{equation}
It appears that the BR tensor squared is quartic in the curvature, which can be originated from a superstring theory as claimed by Ketov in Ref. \cite{Ketov}. Interestingly, the above action can be rewritten in a more convenience form of topological Gauss-Bonnet ${\cal G}$ and Pontryagin $P_4$ densities  such as  \cite{Ketov1,Deser:1999jw}
\begin{equation} \label{action}
 S_{\rm SBR}= \frac{M_p^2}{2}\int d^4 x \sqrt{-g}  \left[R+\alpha_1 R^2 +\alpha_2 \left({\cal G}^2-{P_4}^2\right)\right],
\end{equation}
where 
\begin{equation}
{ \cal G} \equiv  R^2 -4R_{\mu\nu} R^{\mu\nu} +R_{\mu\nu\rho\sigma}R^{\mu\nu\rho\sigma},
\end{equation}
\begin{equation}
{P_4}\equiv \frac{1}{2}\sqrt{-g} \epsilon_{\mu\nu\rho\sigma} R^{\rho \sigma}{}_{\alpha \beta} R^{\mu\nu\alpha\beta}.
\end{equation}
Here, $\epsilon_{\mu\nu\rho\sigma}$ is the Levi-Civita tensor, which is totally antisymmetric with $\epsilon_{0123}=1$. In addition, we have introduced new parameters such as $\alpha_1 \equiv {1}/{(6m^2)}>0$ and $\alpha_2 \equiv { \beta}/{(32m^6)}$ just for convenience.
\subsection{Friedmann-Lemaitre-Robertson-Walker spacetime}
In order to seek de Sitter inflationary solutions, we will consider the spatially flat Friedmann-Lemaitre-Robertson-Walker (FLRW) metric, which is an unique spacetime compatible with the cosmological principle due to its spatial homogeneity and isotropy,
\begin{equation}
ds^2 =-N^2(t)dt^2 +e^{2\alpha(t)} \left(dx^2 +dy^2 +dz^2 \right),
\end{equation}
where $N(t)$ is a lapse function, $\alpha(t)$ is associated with the scale factor, and $t$ is the cosmic time. It turns out that the homogeneity property of FLRW metric will be ensured if both $N$ and $\alpha$ depend only on $t$. It should be noted that $\alpha(t) >0$ is  for any expanding universes; while $\alpha(t) \gg 1$ is for any inflationary universes. As a result, the corresponding Gauss-Bonnet  and  Pontryagin densities are defined to be
\begin{eqnarray}
&&{\cal G}=-\frac{24}{N^5} \dot\alpha^2 \left[ \dot N \dot \alpha - N \left(\ddot\alpha+\dot\alpha^2 \right) \right], \\
&& {P_4}=0.
\end{eqnarray}
 In addition, we have associated results, 
 \begin{equation}
 \sqrt{-g}=N e^{3\alpha}, \quad R= -\frac{6}{N^{2}} \left(\frac{ \dot N}{N} \dot\alpha -\ddot\alpha-2\dot\alpha^2 \right),
 \end{equation}
  which will be useful to define the field equations. In the above expressions, it appears that $\dot\alpha \equiv d\alpha/dt$ and $\ddot\alpha \equiv d^2 \alpha /dt^2$, and so on. 
\subsection{Field equations}
Given the FLRW metric, we are now able to work out the corresponding field equations. As a result, the Euler-Lagrange (EL) equations for $N$ and $\alpha$ are defined to be
\begin{equation}
\frac{\partial {\cal L}}{\partial N} -\frac{d}{dt} \left(\frac{\partial {\cal L}}{\partial \dot N}\right)=0
\end{equation}
and
\begin{equation}
\frac{\partial {\cal L}}{\partial \alpha} -\frac{d}{dt} \left(\frac{\partial {\cal L}}{\partial \dot \alpha}\right) + \frac{d^2}{dt^2} \left(\frac{\partial {\cal L}}{\partial \ddot\alpha}\right)=0, 
\end{equation}
respectively.  Here ${\cal L}$ is nothing but the Lagrangian of the SBR model, whose formalism is given by
\begin{equation}
{\cal L} = \sqrt{-g} \left[R+\alpha_1 R^2 +\alpha_2 \left({\cal G}^2-{P_4}^2\right)\right].
\end{equation}
As a result, these EL equations  will be reduced to
\begin{equation}
\label{eq-1}
6\alpha_1 \left( 2 \dot\alpha \alpha^{(3)} -\ddot\alpha^2 +6 \dot\alpha^2 \ddot\alpha \right) + 96 \alpha_2 \dot\alpha^4 \left(2 \dot\alpha \alpha^{(3)} +3 \ddot\alpha^2 +6\dot\alpha^2 \ddot\alpha - \dot\alpha^4 \right) + \dot\alpha^2 =0
\end{equation}
and 
\begin{eqnarray}
\label{eq-2}
&&6 \alpha_1 \left( 2 \alpha^{(4)} + 12\dot\alpha \alpha^{(3)} +9 \ddot\alpha^2 +18 \dot\alpha^2 \ddot\alpha \right)  +96 \alpha_2  \dot\alpha^2 \left[ 2 \dot\alpha^2 \alpha^{(4)}  + \dot\alpha  \left( 16 \ddot\alpha + 12 \dot\alpha^2 \right) \alpha^{(3)} +12 \ddot\alpha^3  \right. \nonumber\\
&&\left.+45 \dot\alpha^2 \ddot\alpha^2+10 \dot\alpha^4 \ddot\alpha - 3 \dot\alpha^6  \right] +2 \ddot\alpha +3 \dot\alpha^2 =0,
\end{eqnarray}
respectively, once the limit $N(t)\to1$ is taken. Here, $\alpha^{(n)} \equiv d^n \alpha/dt^n$ as higher-order derivatives.  It appears that Eq. (\ref{eq-1}) is a third-order differential equation and can be called the Friedmann equation acting as a constraint equation; while Eq. (\ref{eq-2}) is a fourth-order differential equation describing the evolution of $\alpha$. Due to this property,  the SBR gravity can be regarded as a fourth-order gravity model from now on.
\subsection{Exact de Sitter inflationary solution}
Following Refs. \cite{barrow1,barrow2}, we will take an ansatz, which is associated with the de Sitter solution, for the scale factor such as
\begin{equation}
  \alpha =\zeta t,
  \end{equation}
 by which both Eqs. (\ref{eq-1}) and (\ref{eq-2}) will reduce to the same algebraic equation,
 \begin{equation}
96 \alpha_2  \zeta^6 -1 =0.
\end{equation}
As a result, a non-trivial solution of $\zeta$ can be defined from this equation to be
\begin{equation} 
\zeta =\left(96\alpha_2\right)^{-\frac{1}{6}}.
\end{equation}
As a result, the inflation constraint that $\zeta \gg 1$ will only be satisfied if $0<\alpha_2 \ll 1$. Additionally, all negative values of $\alpha_2$ will correspond to imaginary values of  $\zeta$, which are indeed unphysical. Before ending this section, we would like to note that the Starobinsky term $R^2$ contributes nothing to the found de Sitter inflationary solution. However, its non-trivial role will be highlighted in the stability issue of this solution presented in the next section. 
\section{Stability of the de Sitter inflationary solution}
This section will be used to present a stability analysis based on the powerful dynamical system  method for the obtained de Sitter inflationary solution. According to the standard procedure, e.g., see Refs. \cite{barrow1,barrow2} for more details, we first have to define the corresponding dynamical system of the SBR gravity by introducing suitable dynamical variables,
 \begin{eqnarray}
 &&B=\frac{1}{\dot\alpha^2},\\
&& Q=\frac{\ddot\alpha}{\dot\alpha^2},\\
 && Q_2 =\frac{\alpha^{(3)}}{\dot\alpha^3},
 \end{eqnarray}
 which are dimensionless. Here, $H\equiv \dot\alpha$ is the Hubble parameter. 
 As a result, the corresponding dynamical system turns out to be
 \begin{eqnarray}
  \label{Dyn-1}
 &&B' = -2QB,\\
  \label{Dyn-2}
&& Q' =Q_2 -2Q^2,\\
 \label{Dyn-3}
 &&Q_2 '= \frac{\alpha^{(4)}}{\dot\alpha^4}-3Q Q_2,
 \end{eqnarray}
 where the prime notation stands for a derivative w.r.t. a dynamical time defined as $\tau \equiv \int \dot\alpha dt$. It is noted that the term ${\alpha^{(4)}}/{\dot\alpha^4}$ shown in Eq. (\ref{Dyn-3}) can be figured out from the field equation (\ref{eq-2}), which can be rewritten in terms of the introduced dynamical variables as
\begin{eqnarray}
\label{Dyn-4}
&& 6 \alpha_1B^2 \left( 2 \frac{\alpha^{(4)}}{\dot\alpha^4} + 12Q_2+9 Q^2  +18 Q \right)  +96 \alpha_2   \left[ 2 \frac{ \alpha^{(4)} }{\dot\alpha^4} + Q_2 \left( 16Q + 12  \right)  +12 Q^3 +45 Q^2 +10 Q - 3  \right] \nonumber\\
&& +2 B^3 Q  +3 B^3 =0.
\end{eqnarray}
Mathematically, isotropic fixed points of this dynamical system will be solved from a set of equations given by
 \begin{equation}
 B'=Q'=Q_2' =0.
 \end{equation}
 As a result, these equations will admit a non-trivial solution,
 \begin{equation}
 Q=Q_2 =0
 \end{equation}
  along with an equation of $B$,
 \begin{equation} \label{eq-3}
   \quad B^3 -96 \alpha_2 =0.
 \end{equation}
 It is noted that this fixed point does satisfy the constraint equation coming from Eq. (\ref{eq-1}) as
 \begin{equation}
 6\alpha_1 B^2 \left( 2Q_2 -Q^2 +6Q \right) +96 \alpha_2 \left( 2Q_2 +3Q^2 +6Q-1\right) +B^3=0.
 \end{equation}
 It now becomes clear that the solution of Eq. (\ref{eq-3}), 
 \begin{equation}
 B = \left(96 \alpha_2 \right)^{\frac{1}{3}},
 \end{equation}
 is nothing but the found de Sitter solution due to an important relation that $B =\zeta^{-2}$.
 
 Next, an important issue must be considered is that the stability of the isotropic fixed point, by which we can determine whether the obtained de Sitter inflationary solution is stable or not. In order to do this task, we will first perturb the dynamical system around its isotropic fixed point by taking field perturbations as follows
 \begin{equation}
 B \to B +\delta B, \quad Q \to Q+\delta Q, \quad Q_2 \to Q_2 +\delta Q_2,
 \end{equation}
 where the value of $B$, $Q$, and $Q_2$ is nothing but that of the isotropic fixed point found above. As a result, perturbed autonomous equations can be defined to be
 \begin{eqnarray}
 \label{per-Dyn-1}
 &&\delta B' = -2\left[(\delta Q) B+Q \delta B \right] =-2 B \delta Q ,\\
  \label{per-Dyn-2}
&& \delta Q' =\delta Q_2 -4Q\delta Q = \delta Q_2,\\
 \label{per-Dyn-3}
 &&\delta Q_2 '= \delta \left(\frac{\alpha^{(4)}}{\dot\alpha^4} \right)-3 \left[(\delta Q) Q_2 +Q \delta Q_2 \right]= \delta \left(\frac{\alpha^{(4)}}{\dot\alpha^4} \right),
 \end{eqnarray}
 here the undetermined term, $\delta \left(\frac{\alpha^{(4)}}{\dot\alpha^4} \right)$, will be solved from two perturbed equations coming from Eqs. (\ref{eq-1}) and (\ref{eq-2}) to be
  \begin{equation}
 \delta \left(\frac{\alpha^{(4)}}{\dot\alpha^4} \right) = \frac{1}{6} \left[ \left(384\alpha_2-B^3 \right)\delta Q +18 \left(16\alpha_2-\alpha_1 B^2 \right) \delta Q_2 \right] \left(16\alpha_2+ \alpha_1 B^2\right)^{-1}  .
 \end{equation}
 As a result, taking exponential perturbations,
 \begin{eqnarray}
 && \delta B = A_1 \exp[\mu\tau],\\
  && \delta Q =A_2 \exp[\mu\tau], \\
&& \delta Q_2 =A_3 \exp[\mu\tau],
  \end{eqnarray}
 will lead us to the corresponding matrix equation,
  \begin{equation} \label{stability-equation}
 {\cal M}\left( {\begin{array}{*{20}c}
   A_1  \\
   A_{2}  \\
   A_{3} \\
 \end{array} } \right) \equiv \left[ {\begin{array}{*{20}c}
   {-\mu} & {-2B} & {0 }   \\
   { 0} & {-\mu} & {1 }  \\
     {0  } & { \frac{384\alpha_2- B^3}{6\left(16\alpha_2+\alpha_1 B^2\right)}} & { \frac{3\left(16\alpha_2-\alpha_1 B^2\right)}{16\alpha_2+\alpha_1 B^2}-\mu }  \\
 \end{array} } \right]  \left( {\begin{array}{*{20}c}
    A_1  \\
   A_{2}  \\
   A_{3} \\
 \end{array} } \right) = 0.
\end{equation}
It is well known that non-trivial solutions to this homogeneous system of linear equations of $A_i$ ($i=1-3$) exist only when 
\begin{equation}
\det {\cal M} =0,
\end{equation}
which can be explicitly expanded to be an  equation of $\mu$  given by 
\begin{equation} \label{equation-of-mu}
 \left(16 \alpha_2 +\alpha_1 B^2 \right) \mu^2 - 3 \left(16 \alpha_2 - \alpha_1 B^2   \right) \mu -48\alpha_2 =0,
\end{equation}
where the solution $B^3 =96\alpha_2$ has been used. 
Mathematically, any positive value of $\mu$ will lead to the blowing up of field perturbations as $\tau$ becomes large.  Consequently, this will destroy the stability of the isotropic fixed point. Otherwise, any negative value of $\mu$ will lead to the vanishing of field perturbations as $\tau$ significantly increases. Consequently, this will ensure the stability of the isotropic fixed point. By examining the coefficients of the quadratic equation of $\mu $ shown in Eq. (\ref{equation-of-mu}), it becomes that  the positivity of $\alpha_1$ as required in the pure Starobinsky gravity model will make the isotropic fixed point and therefore the corresponding de Sitter inflationary solution unstable. This is due to an important observation that Eq. (\ref{equation-of-mu}) with $a_0 \equiv -48\alpha_2 <0$ and $a_2 \equiv 16 \alpha_2 +\alpha_1 B^2>0$ will admit at least one positive root $\mu >0$.
Interestingly, the corresponding de Sitter inflationary solution will be stable if $\alpha_1 < -\left(4\alpha_2/9\right)^{1/3}$, or equivalently $a_2 <0$. However, it should be noted that any negative value of $\alpha_1$ is not suitable for the pure Starobinsky gravity model. This means that the SBR gravity model having a negative $\alpha_1$ will not reduce to the pure Starobinsky gravity model having a positive $\alpha_1$ in the limit $\alpha_2 \to 0$.
\section{Conclusions}
We have summarized main results of our study on the de Sitter inflationary solution within the SBR gravity model published recently in Ref. \cite{Do:2023yvg}. In particular, we have shown that the de Sitter inflationary solution can be obtained within the SBR gravity model for $0<\alpha_2\ll 1$. Remarkably, although the coefficient $\alpha_1$ of the quadratic term (or the Starobinsky term) $R^2$ contributes nothing to the value of the obtained inflationary solution, it definitely affects on the stability of this solution. It is important to note that any gravity model admitting unstable de Sitter inflationary solutions will have a possibility to be a realistic inflationary model \cite{Vernov:2021hxo}. A reason for this claim is based on the well-known graceful exit problem, which seems to be unsolved for any stable de Sitter inflationary solutions forming an eternal inflationary phase of the early universe. In this case, quasi-de Sitter solutions will be expected to exist \cite{Ketov1,Vernov:2021hxo}. Similar investigations can be applied to other higher-order gravities. 
\section*{Acknowledgments}
 This work is funded by Vietnam National Foundation for Science and Technology Development (NAFOSTED) under grant number 103.01-2023.50. The author would like to thank Duy H. Nguyen and Tuyen M. Pham very much for their useful help. 

\end{document}